\documentclass{article}%
\usepackage{amsmath}
\usepackage{graphicx}

\newcommand{\PaperTitle}[1]{%
   \begin{center}%
      \begin{large}%
         \textbf {#1} \\%
      \end{large}%
   \end{center}%
}%
\newcommand{\AuthorList}[1]{%
   \begin{center}%
      {#1} \\%
   \end{center}%
}%
\newcommand{\AuthorAffiliations}[1]{%
   \begin{center}%
      {#1}%
   \end{center}%
}%
\newcommand{\SubjectCategories}[1]{%
   \begin{center}%
      Subject Categories: {#1}%
   \end{center}%
}%
\begin{document}%
\PaperTitle{Extending phase field models of grain boundaries to three
  dimensions} 
%
%
\AuthorList{Ryo Kobayashi$^1$, James A. Warren$^2$}
\AuthorAffiliations{$^1$Department of Mathematical and Life 
Sciences, Hiroshima University, Higashi-Hiroshima, 739-8526 Japan\\
  $^2$Center for Theoretical and Computational Materials
  Science and Metallurgy Division\\National Institute of Standards and
  Technology, Gaithersburg, MD 20899, USA} 
\SubjectCategories{Modeling and Simulation, Molten Metal and
  Solidification, Fundamentals}
\section{Abstract}
In this letter we describe a method of extending an existing phase
field model of 
polycrystalline solidification from two to three dimensions (3D).
\section{Introduction and Preliminaries}%
There have been a number of approaches to the modeling of grain
boundaries, all of which have limitations and advantages.  Of
particular interest are phase field models, which have gained
popularity as their ability to compute realistic microstructures
has been demonstrated.  For an overview of this approach, the reader
is recommended some of the review articles on this topic. \cite{GranasyPW04}

A basic model of grain boundaries in 2D (see  \cite{WKLC,kwc})
can be derived from the total free energy 
\begin{eqnarray}
  \label{eq:free_energy_old}
    {\mathcal{F}} =\int dV \big[f(\phi,T) +
     \frac{\alpha^2}{2}|\nabla\phi|^2\nonumber +s g(\phi)|\nabla\theta|+
     \frac{\epsilon^2}{2} h(\phi) |\nabla\theta|^2\big],
\end{eqnarray}
where $f(\phi,T)+\frac{\alpha^2}{2}|\nabla\phi|^2$ are the terms
found in   classical phase field models of solidification, namely the bulk
free energy density, which depends on the phase field $\phi$ and the
temperature $T$, with minima in the liquid 
and solid phases $\phi=0, 1$ plus a  gradient penalty for
interfaces. For this discussion we have omitted terms accounting for
interface energy anisotropy, although such effects are both important
and can be accounted for with well known extensions to this
theory. \cite{WKLC} 

The final terms in the free energy are functions of the gradient in
the orientation, $\theta$; introduced to allow for grain
boundary energy 
misorientation penalties. These terms inclusion also provides a
realistic description of related phenomena such as 
polycrystalline growth and nucleation.\cite{GranasyPBWD04},\cite{Granasy1} The
couplings $g(\phi)$ and $h(\phi)$ are chosen so there are no energy
penalties in the liquid (i.e. $g(0)=h(0)=0$.)  The dynamics of the system
are found by imposing the thermodynamic requirement that $\phi$ and
$\theta$ evolve so as to  minimize the free energy $F$. 

\section{Extension to 3D: Formulation and Solution}
We note that a single angle cannot represent an
orientation in 3D, and thus this concept must be replaced with a more
robust mathematical description of orientation.
Specifically, $\theta$ must be replaced with an object
that captures the three rotational degrees of freedom available in 3D.
Additionally, we must also define the norm of this object.  
With these two mathematical concepts the transition to three 
dimensions is fully posed.
\subsection{Formulation}
There are many ways to represent orientations in 3D,
most quite familiar to crystallographers: Euler angles, rotation
vectors, Rodrigues vectors, quaternions, etc.  All of these
representations are mathematically 
equivalent representations of the group  $SO(3)$ (special orthogonal
group 3D), but retain advantages and disadvantages, depending on the
application.   If we call a member of this group $P$,  then $P$ is a $3
\times 3$ 
orthogonal matrix ($P^T P=I$,where $I$ is the 
identity matrix), and $P$ has a positive determinant $\det P=1$. Thus,
we say $SO(3)$ is naturally embedded in $\mbox{\boldmath$R$}^{9}$, as it can be
represented as a nine-dimensional object (with 6  constraints
originating from the orthogonality condition).

To proceed, we must find the 3D analog to the fundamentally 2D
quantity  $|\nabla\theta|$. A gradient is simply a difference over an
infinitesimal distance, thus we  need to compute the {\sl norm
  of the difference,   in some sense, between two 3D orientations.} 
We consider two possible choices for this measure, employing
a function of two $SO(3)$ matrices $\rho(P,Q)$:
\begin{eqnarray*}
        \mbox{Type I} &:&\hspace{2mm} \rho(P, Q) = | P - Q | = |PQ^{-1} - I |\\
        \mbox{Type II} &:&\hspace{2mm} \rho(P, Q) 
        = \sqrt{2} \cos^{-1}\frac{\mbox{tr}{PQ^{-1}} - 1}{2}
\end{eqnarray*}
Note that, $PQ^{-1}$ is the {\sl misorientation} between two
crystals. The meaning of the Type I measure is trivial, as it measures the  
distance between two matrices in $\mbox{\boldmath$R$}^9$.
The Type II measure, on the other hand, measures the length of the 
geodesic in $SO(3)$ connecting two matrices.  
These two measures coincide when $P$ and $Q$ are infinitesimally
close, but they will yield different values when there is a
discontinuity between $P,Q$ {\sl  as is   often the case for discrete
  computations on a   lattice} 

With these definitions we can now write down our model, by simply
substituting $|\nabla\theta|\rightarrow|\nabla P|$ in
Eqn. \ref{eq:free_energy_old}: 
\begin{eqnarray*}
{\mathcal{F}}= \int dV\big[
f(\phi, T)+\frac{\alpha^{2}}{2}|\nabla \phi|^{2}\nonumber +s g(\phi) |\nabla P| + 
\frac{\epsilon^{2}}{2}h(\phi) |\nabla P|^{2}\big],
\end{eqnarray*}
A more explicit form can be obtained using $|\nabla P| = \sqrt{|\nabla P|^{2}} 
= \sqrt{\sum_{i,j =1}^{3}|\nabla p_{i,j}|^{2}}$, where
$p_{i,j}=[P]_{i,j}$.

\subsection{Solution}
Having posed the above free energy, we must now perform a
minimization, to derive equations of motion for both the phase field
$\phi$ and the orientation.  We must
proceed with care to ensure that the equations of motions keep the
variables in $SO(3)$.  There are several ways to proceed: (1) Derive
equations for 
the constrained free energy, which has only 3 degrees of freedom or
(2) derive equations on $\mbox{\boldmath$R$}^9$, and project the results back
into $SO(3)$. For both methods we need to derive the variational derivative of the
free energy with respect to orientation, which is simply
\begin{eqnarray*}
\frac{\delta \mathcal{F}}{\delta P} = 
- \nabla \cdot \left(sg(\phi) \frac{\nabla P}{|\nabla
    P|}+\epsilon^2 h(\phi)\nabla P \right).
\end{eqnarray*}

In deriving the equations of motion for the
constrained free energy, an element of $SO(3)$ is written in the form  
$P = P(u, v, w)$ where the triplet $(u, v, w)$ is some local
coordinate, for example,  
the Rodrigues vector. The equations of motion for these variables
should be
\begin{equation}
       \tau_u \frac{\partial u}{\partial t} = \left< -\frac{\delta \mathcal{F}}{\delta P},
          \frac{\partial P}{\partial u}\right>
    \label{eq:Diffusion_local_coord}
\end{equation} 
with identical equations for $u\rightarrow v,w$.
Note that the quantity $\left< \cdot, \cdot \right>$ is the usual
inner product in $\mbox{\boldmath$R$}^{9}$ (a fully contracted matrix product),
and $\tau_u$ is an inverse mobility.

Alternatively, to use a projective formulation, we develop
9 equations of motion in $\mbox{\boldmath$R$}^{9}$ keeping the solution within
$SO(3)$ by taking 
a projection of driving force  
 onto the tangential plane of  $SO(3)$. It is given in the form
$\tau_P{\partial P}/{\partial t}  =  \pi_{P}\left( -{\delta
    \mathcal{F}}/{\delta P} \right)$
where $\pi_{P}$ is the projection operator.  This approach allows for
substantially improved  numerical efficiency.  However, we reserve
discussion of this technique for a later publication.

\begin{figure}[htbp]
  \includegraphics[width=13cm]{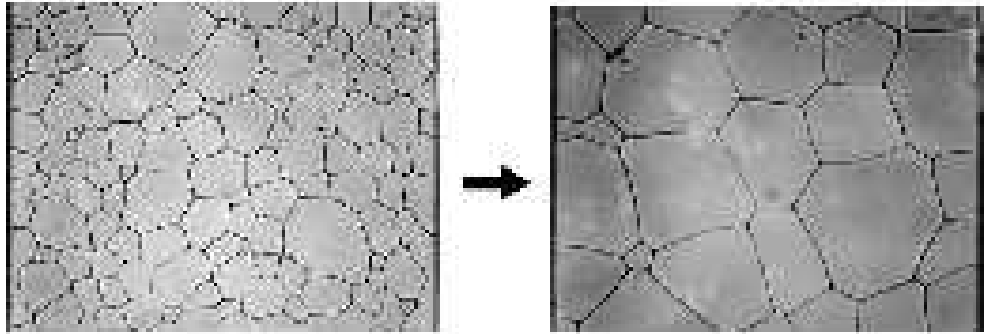}
  \centering
  \caption{Coarsening process of grain structure of succinonitrile.
 The grain structure is almost 2D, while the orientation 
 of each grain is necessarily 3D.
 (courtesy of Drs. Lee and Losert, U. Maryland)}
  \label{fig:Exp_Corsening}
\end{figure}

\begin{figure}[htbp]
  \includegraphics[width=13cm]{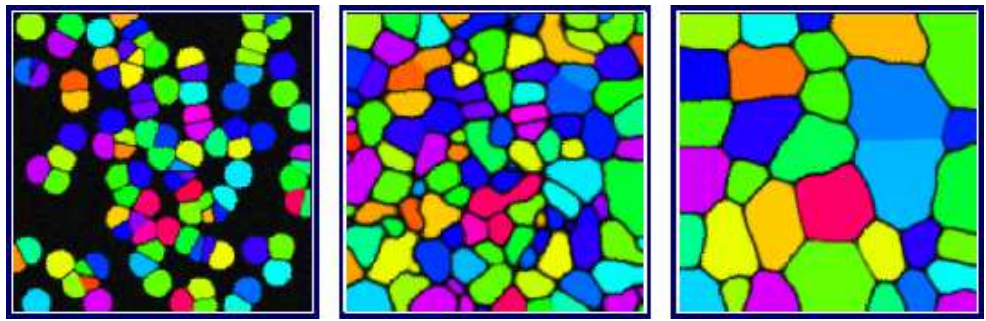}
  \centering
  \caption{Simulation of solidification and coarsening process (the 
  color indicates one of the three Euler angles (all of which were
  solved for).}
  \label{fig:Simulation}
\end{figure}

Following the preceding arguments we can derive the 
evolution equation for $\phi$ and $P$, implement the equations in
computer code and solve.  Our first calculations were for a thin film
where the grains are nearly 2D objects, but 
{\sl their orientation is 3D, and the dynamics will be
  governed by the evolution of all of these angles.}  There are
numerous experimental systems analogous to this calculation (see
Fig.\ref{fig:Exp_Corsening}.) Our results showing growth, impingement
and coarsening are given in Fig. \ref{fig:Simulation}.


Herein, we have extended our previous work in 2D to 3D. For a complete
formulation to be obtained, we must also include the {\sl
  important} consequences of anisotropy. In other words we have
examined the consequences of  misorientation, but not the consequences
of  inclination on the statics and dynamics of grain
boundaries. Additionally, we have yet to account for the underlying
crystal symmetries  These  effects can be included, and will be
discussed in future work.


\begin{thebibliography}{1}

\bibitem{GranasyPW04}
L.~Gr\'an\'asy,T.~Pusztai, and J.~A. Warren, {\em J.  Phys.-Cond. Mat},
vol.~16, pp.~R1205--R1235, 2004; 
W.~J. Boettinger, J.~A. Warren, C.~Beckermann, and A.~Karma, {\em An. Rev. Mat. Res.},  vol.~32, pp.~163--194, 2002;
L.~Q. Chen,  {\em An. Rev. Mat. Res.}, vol.~32,
pp.~113--140, 2002.
\bibitem{WKLC}
J.~A. Warren, R.~Kobayashi, A.~Lobkovsky, and
W.~C. Carter, {\em Acta   Mat.}, vol.~51, pp.~6035--6058, 2003;
\bibitem{kwc}
R.~Kobayashi, J.~A. Warren, and W.~C. Carter,  {\em Physica D}, vol.~140, pp.~141--150, 2000.
\bibitem{GranasyPBWD04}
L.~Gr\'an\'asy,  T.~Pusztai,T.~B\"orzs\"onyi, J.~A. Warren, and J.~F. Douglas,  {\em Nature Mat.},
  vol.~3, pp.~645--650, 2004.
\bibitem{Granasy1}
L.~Gr\'an\'asy, T.~B\"orzs\"onyi, and T.~Pusztai, {\em Phys. Rev. Lett},
  vol.~88, pp.~206105--1--4, 2002.

\end{thebibliography}
\end{document}